# Exact solutions of the field equations for empty space in the Nash gravitational theory

Matthew. T. Aadne[#] and Øyvind G. Grøn[ʘ]

[#] University of Stavanger, P. O. Box 8600 Forus, 4036 Stavanger, Norway

[ʘ] Oslo and Akershus University College of Applied Sciences, Faculty of Technology, Art and Sciences, PB 4 St. Olavs. Pl., NO-0130 Oslo, Norway

**Abstract**

John Nash has proposed a new theory of gravity. We define a Nash-tensor equal to the curvature tensor appearing in the Nash field equations for empty space, and calculate its components for two cases: 1. A static, spherically symmetric space, and 2. The expanding, homogeneous and isotropic space of the Friedmann-Lemaitre-Robertson-Walker (FLRW) universe models. We find the general, exact solution of Nash's field equations for empty space in the static case. The line element turns out to represent the Schwarschild-deSitter spacetime. Also we find the simplest non-trivial solution of the field equations in the cosmological case, which gives the scale factor corresponding to the deSitter spacetime. Hence empty space in the Nash theory corresponds to space with Lorentz Invariant Vacuum Energy (LIVE) in the Einstein theory. This suggests that dark energy may be superfluous according to the Nash theory. We also consider a radiation filled universe model in an effort to find out how energy and matter may be incorporated into the Nash theory. A tentative interpretation of the Nash theory as a unified theory of gravity and electromagnetism leads to a very simple form of the field equations in the presence of matter.

1. **Introduction**

During the last twelve years of his life John Nash tried to develop an alternative theory to Einstein's general theory of relativity. He formulated field equations for empty space, but did not manage to include a description of matter in a way which satisfied his requirements. John Nash presented his work on this theory in lectures [1], but he did not publish any article on it. Neither have we been able to find any analysis by others on the Nash theory.

Nash mentioned several interesting properties of his theory even if the theory was only worked out for empty space. He did among others search for a theory giving a Yukawa-like gravitational field for the static spherically symmetric case, but later dismissed that point of view.

Also Nash mentioned that it is a general consequence of his field equations for empty space that the Ricci scalar obeys a wave equation which reduces to Laplace's equation in the static case. Finally Nash mentioned that his field equations for empty space are satisfied by any solution for empty space of Einstein's field equations with a cosmological constant. This means that phenomena such as accelerated cosmic expansion that comes out of Einstein's theory as an effect of LIVE, causing repulsive gravitation [2],[3], are due to a natural tendency of empty space to expand according to John Nash's theory. This is agreement with the spirit in which Einstein originally interpreted the cosmological constant, before it was reinterpreted by Georges Lemaitre as an expression of the constant energy density of LIVE [4]. Nash did not, however, work out any solution of his field equations.

In the present paper we write out the field equations and find the general solution of the equations for empty static, spherically symmetric space, showing that it is a Schwarzschild de Sitter space time. Furthermore we show that the de Sitter universe solves the Nash equations for empty space with a line element of the FLRW-type.

2. **The field equations for empty space of the Nash theory**

In an unpublished manuscript [1] John Nash has presented a theory of space, time and gravitation. He gave the equations of empty space in the form

$$\Box G^{\mu\nu} + G^{\alpha\beta}\left(2R_{\alpha\phantom{\mu}\beta}^{\phantom{\alpha}\mu\phantom{\beta}\nu} - \frac{1}{2}g^{\mu\nu}R_{\alpha\beta}\right) = 0 . \tag{1}$$

where $\Box$ is the d'Alembertian operator. Defining the Nash curvature tensor

$$N^{\mu\nu} = \Box G^{\mu\nu} + G^{\alpha\beta}\left(2R_{\alpha\phantom{\mu}\beta}^{\phantom{\alpha}\mu\phantom{\beta}\nu} - \frac{1}{2}g^{\mu\nu}R_{\alpha\beta}\right), \tag{2}$$

the Nash field equations for empty space takes the form

$$N^{\mu\nu} = 0 . \tag{3}$$

We shall here present the general solution of these equations for the static spherically symmetric case and a simple, exact solution for expanding, homogeneous and isotropic universe models.

### 3. Static spherically symmetric space.

The line element is described by introducing curvature coordinates where the area of a spherical surface about the origin with a coordinate radius $r$ is $4\pi r^2$. Also, to simplify the calculations we shall restrict ourselves to solutions of the field equations fulfilling the restriction $g_{tt}g_{rr} = -1$. Then the line element takes the form (using units so that the velocity of light is 1)

$$ds^2 = -e^{2\alpha(r)}dt^2 + e^{-2\alpha(r)}dr^2 + r^2 d\Omega^2 .\tag{4}$$

Using the DifferentialGeometry package in Maple we then find the following non-vanishing orthonormal basis components of the Nash tensor,

$$N^{\hat{t}\hat{t}} = 2r^{-4}e^{4\alpha}\left(-r^3\alpha''' - r^2\alpha'' - 6r^3\alpha'\alpha'' + 2r\alpha' - 2r^2\alpha'^2 - 4r^3\alpha'^3 + e^{-2\alpha} - 1\right),$$
$$N^{\hat{r}\hat{r}} = 2r^{-4}e^{4\alpha}\left(r^3\alpha''' + 3r^2\alpha'' + 6r^3\alpha'\alpha'' - 2r\alpha' + 6r^2\alpha'^2 + 4r^3\alpha'^3 + e^{-2\alpha} - 1\right),\tag{5}$$
$$N^{\hat{i}\hat{i}} = r^{-4}e^{4\alpha}[r^4\alpha'''' + 4r^3\alpha''' + 10r^4\alpha'\alpha''' - 2r^2\alpha'' + 32r^2\alpha'\alpha'' + 6r^4\alpha''^2 + 36r^4\alpha'^2\alpha''$$
$$- 4r\alpha' - 8r^2\alpha'^2 + 32r^3\alpha'^3 + 16r^4\alpha'^4 + 2(1+2r\alpha')e^{-2\alpha} - 2].$$

where $' = d/dr$. It follows from these expressions that

$$N^{\hat{t}\hat{t}} + N^{\hat{r}\hat{r}} = 4r^{-4}e^{4\alpha}\left(r^2\alpha'' + 2r^2\alpha'^2 + e^{-2\alpha} - 1\right).\tag{6}$$

Hence the Nash equations for empty space imply that

$$r^2\left(\alpha'' + 2\alpha'^2\right) + e^{-2\alpha} - 1 = 0 ,\tag{7}$$

which may be written

$$\left[r^2 e^{2\alpha}\alpha' + r\left(1 - e^{2\alpha}\right)\right]' = 0 .\tag{8}$$

Integration gives

$$r^2 e^{2\alpha}\alpha' + r\left(1 - e^{2\alpha}\right) = C ,\tag{9}$$

where $C$ is an arbitrary constant. Furthermore

$$r^2 e^{2\alpha}\alpha' + r\left(1 - e^{2\alpha}\right) = -\left(r^4/2\right)\left[\left(1 - e^{2\alpha}\right)/r^2\right]' ,\tag{10}$$

so eq.(9) can be written

$$-\left(r^4/2\right)\left[\left(1 - e^{2\alpha}\right)/r^2\right] = C.\tag{11}$$

Integrating once more leads to

$$e^{2\alpha} = 1 - \frac{3C}{2r} - Dr^2 .\tag{12}$$

With the identification $C = (2/3)r_s$, $D = \Lambda/3$ this is the line element of the Schwarzschild-deSitter spacetime outside a spherical mass with Schwarzschild radius $r_s$ in a universe dominated by LIVE with energy density $\rho_\Lambda$ given in terms of the cosmological constant $\Lambda$ by $\kappa \rho_\Lambda = \Lambda$, where $\kappa = 8\pi G$ is Einstein's gravitational constant.

## 4. Homogeneous and isotropic universe models

The standard form of the line element is in this case

$$ds^2 = -dt^2 + a^2(t)\left(\frac{dr^2}{1-kr^2} + r^2 d\Omega^2\right). \tag{13}$$

For this line-element non-vanishing the components of the Nash-tensor are

$$\begin{aligned} N^{\hat{t}\hat{t}} &= -(3/a^4)\left(2a^2\dot{a}\dddot{a} + 2a\dot{a}^2\ddot{a} - a^2\ddot{a}^2 - 3\dot{a}^4 + k^2 - 2k\dot{a}^2\right), \\ N^{\hat{i}\hat{i}} &= (1/a^4)\left(2a^3\dddot{a} + 4a^2\dot{a}\ddot{a} + 3a^2\ddot{a}^2 - 12a\dot{a}^2\ddot{a} + 3\dot{a}^4 - 4k a\ddot{a} + 2k\dot{a}^2 - k^2\right) \end{aligned} \tag{14}$$

The trace of the Nash-tensor is

$$Tr(N) = (6/a^3)\left(a^2\dddot{a} + 3a\dot{a}\ddot{a} + a\ddot{a}^2 - 5\dot{a}^2\ddot{a} - 2k\ddot{a}\right). \tag{15}$$

We shall here consider flat universe models. The components and trace of the Nash tensor can then be given in terms of the Hubble parameter, $H = \dot{a}/a$, as

$$\begin{aligned} N^{\hat{t}\hat{t}} &= -3\left(2H\ddot{H} + 6H^2\dot{H} - \dot{H}^2\right), \\ N^{\hat{i}\hat{i}} &= 2\dddot{H} + 12H\ddot{H} + 9\dot{H}^2 + 18H^2\dot{H} \end{aligned} \tag{16}$$

and

$$Tr(N) = 6\left(\dddot{H} + 7H\ddot{H} + 12H^2\dot{H} + 4\dot{H}^2\right). \tag{17}$$

A further simplification of the expression for $N^{\hat{t}\hat{t}}$ is obtained by introducing $y(t) = \sqrt{H}$. Then the components of the Nash tensor take the form

$$\begin{aligned} N^{\hat{t}\hat{t}} &= -12y^3\left(\ddot{y} + 3y^2\dot{y}\right) \\ N^{\hat{i}\hat{i}} &= 4\left(y\dddot{y} + 6y^3\ddot{y} + 3\dot{y}\ddot{y} + 9y^5\dot{y} + 15y^2\dot{y}^2\right) \end{aligned} \tag{18}$$

The Nash equations for empty space reduce to

$$\ddot{y} + 3y^2\dot{y} = 0, \tag{19}$$

$$y\dddot{y} + 6y^3\ddot{y} + 3\dot{y}\ddot{y} + 9y^5\dot{y} + 15y^2\dot{y}^2 = 0. \tag{20}$$

Equatens (19) and (20) may be written, respectively, as

$$\left(\dot{y} + y^3\right)^\cdot = 0, \tag{21}$$

and

$$y(\dot{y}+y^3)^{\cdot\cdot} + (3/2)\left[(\dot{y}+y^3)^2\right]^{\cdot} = 0. \tag{22}$$

Hence, eq.(21) implies that eq.(22) is fulfilled, so there is only one independent field equation for this problem. The general solution is thus found by solving eq.(19).

We shall integrate equation (19) in two steps in order to discuss an interesting special case. A first integration gives

$$\dot{y} + y^3 = C_1, \tag{23}$$

where $C_1$ is a constant of integration. Let us consider first the special case $C_1 = 0$. Integrating eq.(23) with $\lim_{t \to 0} H(t) = \infty$ and $a(t_0) = 1$ where $t_0$ is the present age of the universe, then gives

$$a(t) = t/t_0, \tag{24}$$

which is the scale factor of the Milne universe, i.e. of the Minkowski spacetime as described in an expanding reference frame.

We then consider the case $C_1 \neq 0$. A particular solution of eq.(23) is then $H = C_1^{2/3}$, $\dot{y} = 0$. In this case the Hubble parameter is constant. Integrating once more with $a(t_0) = 1$ gives

$$a(t) = e^{H(t-t_0)}. \tag{25}$$

This is the scale factor of the deSitter universe, which in Einstein's theory is a spacetime dominated by LIVE with a constant density that may be represented by a cosmological constant.

Eq.(23) is an Abel differential equation of the first kind. The general solution of this equation can only be given implicitly with $t$ as a function of $y$. This solution does not appear in Einstein's theory.

### 5. How can energy-momentum be represented in the Nash field equations?

The energy-momentum density tensor of electromagnetic radiation is trace free. We shall therefore start by considering the equation $Tr(N) = 0$. From eq.(17) this takes the form

$$\dddot{H} + 7H\ddot{H} + 12H^2\dot{H} + 4\dot{H}^2 = 0. \tag{26}$$

which may be written

$$(\dot{H} + 2H^2)^{\cdot\cdot} + 3H(\dot{H} + 2H^2)^{\cdot} = 0. \tag{27}$$

We shall not try to find the general solution of this equation, but restrict ourselves to investigate a class of solutions obeying

$$(\dot{H} + 2H^2)^{\cdot} = 0. \tag{28}$$

Integration gives

$$\dot{H}+2H^2=2H_0^2,\qquad(29)$$

where $H_0$ is a constant. With $H_0=0$ and the usual normalization of the scale factor, $a(t_0)=1$, integration of eq.(29) gives

$$a(t)=\sqrt{1+2(t-t_0)}.\qquad(30)$$

This has the same form as the scale factor of a flat radiation dominated FLRW-universe in the Einstein theory.

Inserting this scale factor into eq.(14) we find that for this solution the components of the Nash tensor are

$$N^{\hat{t}\hat{t}}=0\ ,\ N^{\hat{i}\hat{i}}=30/a^{12}.\qquad(31)$$

Hence, the field equations cannot have the form that the Nash tensor is proportional to the energy-momentum density tensor.

A particular solution of eq.(29) with $H_0\neq 0$ is $H=H_0$, which gives the scale factor (25) representing the deSitter universe. The general solution of eq.(29) with $H_0\neq 0$ is

$$H(t)=(H_0/2)\cotanh\left[H_0(t-C_3)\right].\qquad(32)$$

Integrating once more with $a(0)=0$ and $C_3=0$ gives

$$a(t)=a_1\sinh^{1/2}(H_0 t).\qquad(33)$$

Solving the Friedmann equations of Einstein's theory with a non-vanishing cosmological constant for a flat FLRW-universe filled by a perfect fluid with equation of state $p=w\rho$ and present density $\rho_0$ gives [5]

$$a(t)=\left(\frac{3\kappa\rho_0}{\Lambda}\right)^{\frac{1}{3(1+w)}}\sinh^{\frac{2}{3(1+w)}}\left[\frac{(1+w)\sqrt{3\Lambda}}{2}t\right],\qquad(34)$$

For electromagnetic radiation $w=1/3$, and eq.(33) becomes

$$a(t)=\left(\frac{3\kappa\rho_0}{\Lambda}\right)^{\frac{1}{4}}\sinh^{\frac{1}{2}}\left(2\sqrt{\frac{\Lambda}{3}}t\right).\qquad(35)$$

Hence with the identification $a_1=(3\kappa\rho_0/\Lambda)^{1/4}$, $H_0=2\sqrt{\Lambda/3}$ the scale factor (33) is similar to the scale factor (35) of a flat FLRW-universe with radiation and LIVE in the Einstein theory. Inserting this solution in eq.(14) gives the following components of the Nash-tensor,

$$N^{\tilde{i}\tilde{i}} = (1/3)N^{\tilde{t}\tilde{t}} = -\frac{H_0^4}{4\sinh^2(H_0 t)}. \tag{36}$$

This shows that if we think of the Nash theory as a theory of space-time and gravitation, for either an empty universe or a universe filled by radiation, then the field equations may be given the form that the Nash-tensor is proportional to the energy-momentum density tensor with a negative proportionality constant. In this theory there would be no need for dark energy, because the accelerated expansion is according to this theory due to a natural tendency of empty space to expand exponentially.

It is not easy to find a natural extension of the theory to describe matter. In the case of dust, for example, the scale factor of Einstein's theory is found from eq.(34) with $w = 0$. This gives [6]

$$a(t) = \left(\frac{3\kappa\rho_0}{\Lambda}\right)^{\frac{1}{3}} \sinh^{\frac{2}{3}}\left(\frac{\sqrt{3\Lambda}}{2}t\right). \tag{37}$$

With this scale factor the components of the Nash-tensor are

$$N^{\tilde{i}\tilde{i}} = N^{\tilde{t}\tilde{t}} = \frac{4H_0^4}{3\sinh^2(H_0 t)}. \tag{38}$$

If the Nash-tensor is proportional to the energy-momentum tensor with a positive proportionality constant, eq.(38) would represent a Zel'dovich fluid with equation of state $p = \rho$ (not dust). Hence, one may tentatively suggest that the Nash-field equations might have the form $N + \kappa T_{EM} = \kappa T_M$, where $T_{EM}$ is the energy-momentum density tensor of an electromagnetic field (including radiation), and $T_M$ is the corresponding tensor of matter.

6. Conclusion

It has been asked whether John Nash's "interesting equation" really is interesting, and answered that it is a matter of taste, but also that similar equations have interested physicists in the recent past. Nash's field equation results from an action with squared curvature terms of the type

$$\int \sqrt{-g}\left(2R^{\alpha\beta}R_{\alpha\beta} - R^2\right)dx^1 dx^2 dx^3 dx^4$$

It was noted by Nash that "the general class of Lagrangians including ours (written above) has previously caught the attention of theorists. But my observation was, from my reading in the area, that those who had been broadly interested in this general family of Lagrangians had never focused effectively on the specific choice."

It has been pointed out that this class of theories has a problem if one tries to quantize the theory, connected with so-called "ghosts". Such theories result in ghost fields with negative norm, i.e. with negative probabilities, which is clearly unphysical. However, there have been attempts to solve such problems [7], [8].

In the present paper we focus upon investigating classical aspects of Nash's theory. We have defined the Nash tensor to be the curvature tensor appearing in the Nash field equations and calculated its components for a static spherically symmetric space and for the FLRW line-element.

Then we found exact solutions of the Nash field equations and showed that they describe the Schwarzschild-deSitter spacetime in the static case and the deSitter spacetime in the cosmological case. This shows that in the Nash theory space has a natural tendency to have exponentially accelerated expansion, making it unnecessary to introduce dark energy in order to account for the observed accelerated expansion of the universe.

A main unsolved problem is to device a proper form of the field equations in general, i.e. for space containing matter. Motivated by the fact that the energy-momentum tensor of electromagnetic fields is trace free, we have made a preliminary investigation of this problem solving the equation obtained by putting the trace of the Nash tensor equal to zero. In the cosmological case the solution is indeed a line element corresponding to that of a space filled with electromagnetic radiation and LIVE in Einstein's theory. Inserting the scale factor of this solution into the expressions of the components of the Nash tensor we found that they obey $N^{\hat{i}\hat{i}} = (1/3) N^{\hat{t}\hat{t}} < 0$. Hence field equations of the form $N^{\mu\nu} + T_{EM}^{\mu\nu} = 0$ are fulfilled by this solution. Comparing with the corresponding solution of Einstein's theory, the line element has the same form as that of a universe with radiation and LIVE in Einstein's theory.

Inserting the line element of Einstein's theory for a universe with LIVE and dust in Einstein's theory, we found that the components of the Nash tensor obey, $N^{\hat{i}\hat{i}} = N^{\hat{t}\hat{t}} > 0$. This shows that the equations $N + \kappa T_{EM} = \kappa T_M$ are fulfilled by all the solutions we have considered.

If the Nash theory is interpreted as a unified theory of gravitation and electromagnetism, thinking of the electromagnetic energy-momentum density tensor as representing a component of the structure of space, this form of the field equations would be natural. Hence we suggest that the Nash-theory may be conceived of as a unified theory of gravity and electromagnetism where in addition dark energy in the form of LIVE is replaced by a natural tendency of space to have accelerated expansion, making the introduction of dark energy or a cosmological constant superfluous.

**Acknowledgement**

We would like to thank Torkild Jemterud for suggesting to us that we investigate the Nash theory.